\documentclass[prl,aps,twocolumn,showpacs,floatfix]{revtex4}
\input epsf

\usepackage{graphicx}% Include figure files
\usepackage{dcolumn}% Align table columns on decimal point
\usepackage{amssymb}
\usepackage{amsmath}
\usepackage{latexsym}
\usepackage{verbatim}

\begin{document}
\preprint{APS/123-QED}

\title{
The Price of Anarchy in Transportation Networks: Efficiency and
Optimality Control
}

\author{Hyejin Youn}
\affiliation {Department of Physics, Korea Advanced Institute of
Science and Technology, Daejeon 305-701, Korea}

\author{Michael T. Gastner}
\affiliation {Santa Fe Institute, 1399 Hyde Park Road, Santa Fe,
NM 87501, USA} \affiliation {Department of Computer Science,
University of New Mexico, Albuquerque, NM 87131, USA}

\author{Hawoong Jeong}
\email{hjeong@kaist.ac.kr} \affiliation {Department of Physics,
Korea Advanced Institute of Science and Technology, Daejeon
305-701, Korea}

\begin{abstract}
Uncoordinated individuals in human society pursuing their
personally optimal strategies do not always achieve the social
optimum, the most beneficial state to the society as a whole.
%MG(04.08.08): "often" saves a few characters. No comma in front
%  of "which".
%Instead, strategies form Nash equilibria, which are, in general,
Instead, strategies form Nash equilibria which are often
%\MG(04.08.08)
socially suboptimal. Society, therefore, has to pay a \emph{price
%of anarchy} for the lack of coordination among its members which
%is often difficult to quantify in engineering, economics, and
%policymaking. Here we present a quantitative assessment of this
%MG(04.08.08): that it's difficult to quantify is not so central
%  for the content of the paper.
of anarchy} for the lack of coordination among its members. Here
we assess this
%\MG(04.08.08)
%MG(04.08.08): rearranging the position "travel time". By not
%  explicitly mentioning the cities, we save a few characters.
%price of anarchy by analyzing the networks of principal roads in
%Boston, London, and New York, where one's travel time serves as
%the relevant cost to be minimized. Our simulation shows that
price of anarchy by analyzing the travel times in road networks of
several major cities. Our simulation shows that
%\MG(04.08.08)
uncoordinated drivers possibly waste a considerable amount of
%MG(04.08.08): One might argue that the socially optimal traffic
%  is already implied.
%their travel time compared to socially optimal traffic.
their travel time.
%\MG(04.08.08)
Counterintuitively, simply blocking certain streets can partially
%MG(04.08.08): Condensing the last sentence.
%improve the traffic conditions. An analysis of various complex
%networks indicates the possibility of similar paradoxes for
%networked physical systems with equilibrium principles (e.g.,
%Kirchhoff's laws in electric circuits).
improve the traffic conditions. We analyze various complex
networks and discuss the possibility of similar paradoxes in
physics.
%\MG(04.08.08)
\end{abstract}
\maketitle

Many real-world transportation systems in human societies are
characterized by networked structures and complex agents
interacting on these networks~\cite{Kossinets2006}. Understanding
the agents' behaviors has important consequences for the optimal
design and control of, for example, the Internet, peer-to-peer, or
vehicle networks~\cite{Buchanan}. In fact, optimality has long
been a key principle in science. In particular, many branches of
physics are governed by principles of least action or minimum
energy in the same way that maximizing utility functions is
crucial in economics. For example, the flow of currents in a
resistor network can be derived by minimizing the energy
dissipation. One might expect that traffic flows in transportation
networks follow a similar optimization principle.  It is indeed
reasonable to assume that humans opt for the strategies that
maximize their personal utility.  However, this does not mean that
flows in transportation networks minimize the cost for all users
%MG(20.07.08): Well, I'll be chivalrous and go without self-citation.
%HJ(21.07.08): We can keep this
 as is sometimes assumed~\cite{Gastner2006}.
%\HJ(21.07.08): We can keep this
%\MG(20.07.08).
On the contrary, we will demonstrate that the flows can in reality
still be far from optimal even if all individuals search for the
quickest paths and if complete information about the network and other
users' behaviors is available.  Thus, traffic networks can be inherently
inefficient -- a fact rarely investigated in previous work on traffic
flows~\cite{Qiu}.

In this paper, we investigate decentralized transportation
networks where each directed link from node $i$ to $j$ is
associated with a \emph{delay} $l_{ij}$, the time needed to travel
along the link. In most real networks, delays depend noticeably on
the flow~\cite{Levinson1998}, i.e., the number of downloads,
vehicles, etc.\ per unit time. For example, a single vehicle
easily moves at the permitted speed limit on an empty road, yet
slows down if too many vehicles share the same road. Thus, the
choices of some users can cause delays for others and possibly
conflict with everyone's goal to reduce the overall delay in the
network. As a game-theoretic consequence, the best options for
individual users form a Nash equilibrium, not necessarily a social
optimum.

Consider, for instance, the simple network depicted in
Fig.~1(a)~\cite{Roughgarden}. Suppose that there is a constant
flow of travellers $F$ between the nodes $s$ and $t$ which are
connected by two different types of links: a short but narrow
bridge $A$ where the effective speed becomes slower as more cars
travel on it, and a long but broad multi-lane freeway $B$ where
congestion effects are negligible. Suppose the delay on link $A$
is proportional to the flow, $l_{A} ( f_{A} ) = f_{A}$, while the
delay on $B$ is flow-independent, $l_{B} ( f_{B} ) = 10$, where
$f_{A(B)}$ is the flow on link $A(B)$. The total time spent by all
users is given by the ``cost function" $\mathcal{C} ( f_{A} ) =
l_{A} ( f_{A} )\cdot f_{A} + l_{B} ( f_{B} )\cdot f_{B}$ where the
flow on $B$ is equal to $f_{B} = F - f_{A}$. It is easily verified
that $\mathcal{C}$ attains its minimum for $f_{A} = 5$ if the
total flow satisfies $F \geq 5$. If $F = 10$, for example, each
link should be taken by exactly half of the users, resulting in
$\mathcal{C} = 75$ (Fig.~1b).

In this social optimum, every user on link $B$ could reduce his
delay from 10 to 6 by switching paths, which poses a social
dilemma: as individuals, users would like to reduce their own
delays, but this reduction comes at an additional cost to the
entire group. In our example, as long as $l_{A} \neq l_{B}$, there
will be an incentive for the users experiencing longer delays to
shift to another link. If all users decide to put their own
interests first, the flow will be in a Nash equilibrium where no
single user can make any individual gain by changing his own
strategy unilaterally. All users take the link $A$, as shown in
Fig. 1(c), at the total cost of $\mathcal{C}=100$. Experimental
tests indicate that human subjects approach the problem of finding
paths in a network from this latter self-interested perspective,
rather than from the former altruistic point of
view~\cite{Rapoport2005}. This behavior, known as {\it Wardrop's
principle}, is observed even if, as in our example, not a single
user experiences a shorter travel time in this Nash equilibrium
than in the social optimum. Furthermore, if all functions
$l_{ij}(f_{ij})$ are strictly increasing (as in most realistic
cases) and the flows $f_{ij}$ are continuous, one can prove that
there is always exactly one Nash equilibrium~\cite{Roughgarden}.

\begin{figure}
\includegraphics[width=6.5cm]{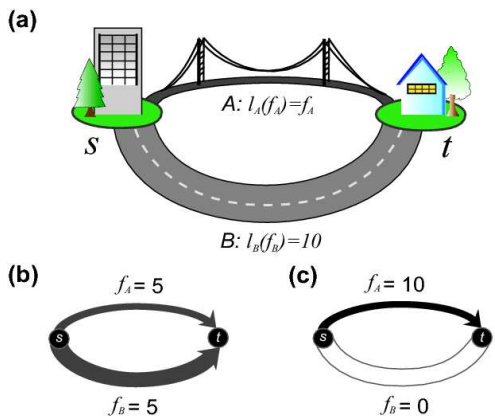}
\caption{(color) Illustration of the price of anarchy. (a) Suppose
$F$ = 10 users travel per unit time from $s$ to $t$. (b) The
socially optimal flow sends five users along each link, thus the
total cost is $\mathcal{C}=75$. (c) In the Nash equilibrium with
$f_A=10$ and $f_B =0$, $\mathcal{C}=100$ is higher than in (b).}
\label{fig:poa}
\end{figure}

Although differences between Nash equilibria and social optima
occur frequently in social science, only few papers have studied
the difference between optimal and actual system performance in
real transportation networks~\cite{Qiu}. To shed light on this
issue, we have analyzed Boston's road network shown in Fig.~2(a).
The 246 directed links in our network are segments of principal
roads, and their intersections form 88 nodes. Delays are assumed
to follow the Bureau of Public Roads (BPR) function widely used in
civil engineering, $ l_{ij} = \frac{d_{ij}}{v_{ij}} \Big[ 1 +
\alpha \Big( \frac{f_{ij}}{p_{ij}} \Big)^{\beta} \Big]$. Here
$d_{ij}$ is the distance of the link between $i$ and $j$, $v_{ij}$
the speed limit (35 mph on all links, for simplicity), $f_{ij}$
the flow, and $p_{ij}$ the capacity of the road segment. The
parameters $\alpha$ and $\beta$ have been fitted to empirical
data~\cite{Singh} as $\alpha = 0.2$ and $\beta= 10$, i.e., the
delays increase very steeply for large traffic volumes. Capacity
is defined as the traffic volume at ``level of service E" which is
approximately 2000 vehicles per hour multiplied by the number of
lanes~\cite{Roess}. We used Google Maps to identify the principal
roads, measure the distances $d_{ij}$, and count the number of
lanes for each direction.

\begin{figure*}
\includegraphics[width=17.8cm]{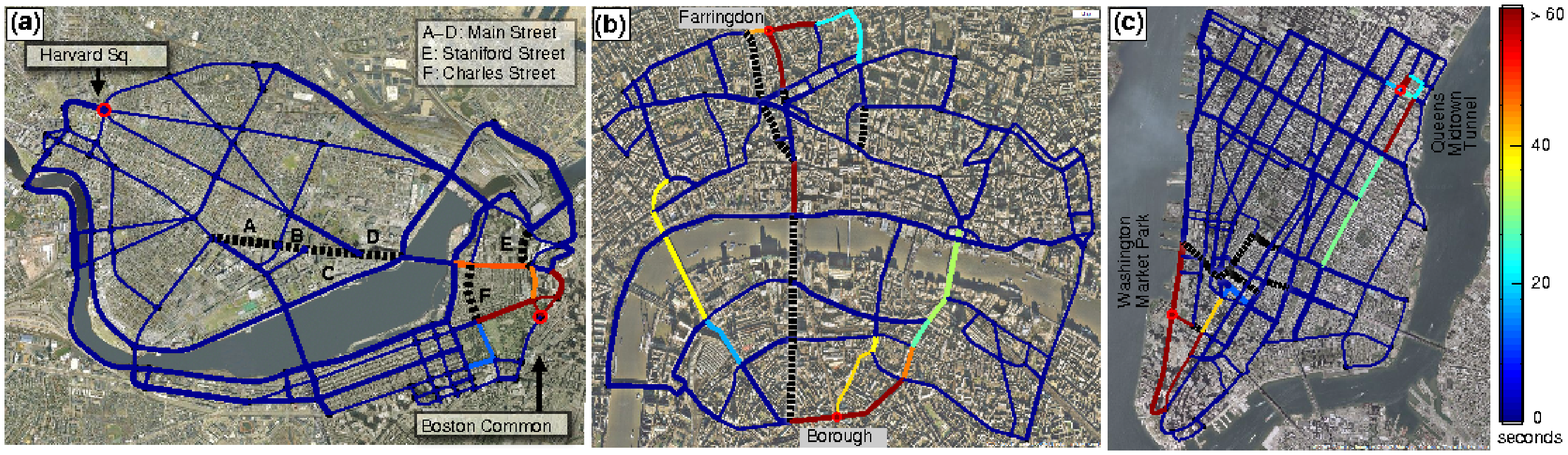}
\caption{(color) Networks of principal roads (both solid and
dotted lines; the thickness represents the number of lanes). (a)
Boston-Cambridge area, (b) London, UK, and (c) New York City. The
color of each link indicates the additional travel time needed in
the Nash equilibrium if that link is cut (blue: no change, red:
more than 60 seconds additional delay). Black dotted lines denote
links whose removal reduces the travel time, i.e., allowing
drivers to use these streets in fact creates additional
congestion. This counter-intuitive phenomenon is called "Braess's
paradox."} \label{fig:networks}
\end{figure*}

Next we have calculated the flows $f_{ij}$ for various total
traffic volumes $F$ from Harvard Square to Boston Common. The
socially optimal flows $f_{ij}^{SO}$ are determined by minimizing
%MG(20.07.08): Well, I guess we have to finally cut it.
%the cost to society per unit time which is the sum over
%everybody's individual delay $ \mathcal{C} = \sum_{link (i,j)}
%l_{ij}(f_{ij})f_{ij}$. This optimization problem, satisfying flow
the cost to society per unit time $ \mathcal{C} = \sum_{link
(i,j)} l_{ij}(f_{ij})f_{ij}$. This optimization problem,
satisfying flow
%\MG(20.07.08)
conservation at each intersection, can be solved with standard
convex minimum cost flow algorithms~\cite{Ahuja1993}. For the Nash
equilibrium, we can use the fact that the equilibrium flows
$f_{ij}^{NE}$ minimize the objective function~\cite{Roughgarden}
$\widetilde{\mathcal{C}}= \sum_{link (i,j)} \int_{0}^{f_{ij}}
l_{ij}(f')df' $. The \emph{price of anarchy} (PoA) is defined as
the ratio of the total cost of the Nash equilibrium to the total
cost of the social optimum~\cite{Papadimitriou} indicating the
inefficiency of decentralization; for example in Fig.~1,
$\text{PoA}=100/75=4/3$, or in general
\begin{equation}
\text{PoA} = \frac{\sum{l_{ij}(f_{ij}^{NE}) \cdot f_{ij}^{NE}
}}{\sum{l_{ij}(f_{ij}^{SO}) \cdot f_{ij}^{SO} }}. \label{PoA}
\end{equation}
4/3 is in fact the upper bound for the PoA in networks with affine
delays, i.e., $\beta=1$~\cite{Roughgarden,Friedman04}.  For larger
$\beta$, the theoretical maximum
%HJ(04.08.08) becomes -> is
is
%\HJ(04.08.08)
higher, but here we are more interested in typical than in
worst-case network topologies. For $\beta=10$, Fig.~3(a) shows the
PoA versus the total traffic volume $F$ for Boston's roads. Except
for very small $F$, the Nash equilibrium cost is higher than the
social optimum so that $\text{PoA} > 1$. The worst ratio occurs
for a traffic volume of $10,000$ vehicles per hour -- a quite
realistic flow, see~\cite{DATA} -- where $\text{PoA} \approx
1.30$, i.e., individuals waste 30\% of their travel time for not
being coordinated.

To what extent are properties of the PoA observed in Boston's road
network characteristic of networks with flow-dependent costs?
Among road networks, the results appear to be typical as suggested
by an analysis of the road networks of London and New York in
Fig.~2. London's network consists of 82 intersections and 217
links marked as principal roads by Google Maps. We find that the
PoA can increase up to 24\% for trips between the Borough and the
Farringdon underground stations (Fig.~3(a) inset). Similar results
also hold for New York, consisting of 125 intersections and 319
streets. The inset of Fig.~3(a) shows that the PoA can be as high
as 28\% when 12,000 vehicles per hour travel from Washington
Market Park to Queens Midtown Tunnel. The results remain qualitatively
similar for different sets of sources and destinations suggesting that
a high PoA can generally become a serious problem.

To gain further theoretical insight, we also constructed four
ensembles of bidirectional model networks with distinct underlying
structures~\cite{ModelNetwork}: a simple one-dimensional lattice
with connections up to the third-nearest neighbors and periodic
boundary conditions, Erd\H{o}s-R\'{e}nyi random graphs with links
between randomly drawn pairs of nodes, small-world networks with a
rewiring probability 0.1, and Barab\'{a}si-Albert networks with
broad degree distributions. All the networks contain 100 nodes and
have an average degree of 6. Every link between nodes $i$ and $j$
has a delay of the form $ l_{ij} = a_{ij}f_{ij} + b_{ij} $, where
$a_{ij} = a_{ji}$ is a random integer equal to 1, 2, or 3, and
$b_{ij} = b_{ji}$ between 1 and 100. This affine cost function
captures essential properties of links in important physical
networks. In electric circuits, for example, the flow $f_{ij}$ is
an electric current and the delay $l_{ij}$ can be interpreted as
the voltage difference between $i$ and $j$. An affine
current-voltage characteristic occurs in circuits with a
combination of Ohmic resistors (resistances $a_{ij}$) and Zener
diodes (breakdown voltages $b_{ij}$). Further examples with affine
cost functions include mechanical, hydraulic, and thermal
networks~\cite{Penchina}.

For each model network, we go through every pair of nodes to
calculate the PoA for various total flows $F$. Then the results
are averaged over 50 networks to find the mean
$\langle$PoA($F$)$\rangle$  for each ensemble as plotted in
Fig.~3(b). After averaging over many pairs, there are no longer
multiple local maxima as in Fig.~3(a). Instead, we find unimodal
functions for all ensembles with a steep increase for small $F$
and a long tail for large flows. The qualitative behavior can be
understood as follows. The social optimum minimizes $ \mathcal{C}
= \sum (a_{ij}{f_{ij}}^2 + b_{ij}f_{ij}) $ whereas the flow in the
Nash equilibrium minimizes $ \widetilde{\mathcal{C}} = \sum (
\frac12 a_{ij}{f_{ij}}^2 + b_{ij}f_{ij}) $. In the limit $F
\rightarrow 0$, both objective functions become identical and,
therefore, $\langle$PoA$\rangle$ $\rightarrow 1$. For $F
\rightarrow \infty$, the quadratic terms in the sums dominate,
hence $\mathcal{C}/\widetilde{\mathcal{C}} \rightarrow 2$, i.e.,
both objective functions are minimized by the same asymptotic flow
pattern $f_{ij} /F$ and $\langle$PoA$\rangle$ again approaches 1.
The maximum $\langle$PoA$\rangle$ occurs roughly where the
quadratic and linear terms in the objective functions are
comparable, i.e., $a_{ij} f_{ij} \approx b_{ij}$ for paths with
positive flow. Ignoring correlations between $a_{ij}$ and
$f_{ij}$, we have $\langle f_{ij} \rangle \approx \langle b_{ij}
\rangle / \langle a_{ij} \rangle$. Since $F = c \langle f_{ij}
\rangle$ where $c$ is a factor bigger than but of the order of 1,
we estimate the maximum $\langle$PoA$\rangle$ to be at $F_{max}
\approx c \langle b_{ij} \rangle / \langle a_{ij} \rangle$. In our
example, $\langle a_{ij} \rangle = 2$ and $\langle b_{ij} \rangle
= 50.5$, so we predict $F_{max}$ to be bigger than but of the
order of 25. Numerically, we find the maxima for our four
ensembles to be between 30 and 60 in good agreement with our
estimate. Barab\'{a}si-Albert networks tend to have the lowest
$\langle$PoA$(F)\rangle$ and small-world networks the highest, but
the statistical dependence between $\langle$PoA$\rangle$ and $F$
is strikingly similar among all ensembles.

\begin{figure}
%MG(21.07.08): The original size seemed too big. 7cm look just fine.
\includegraphics[width=7cm]{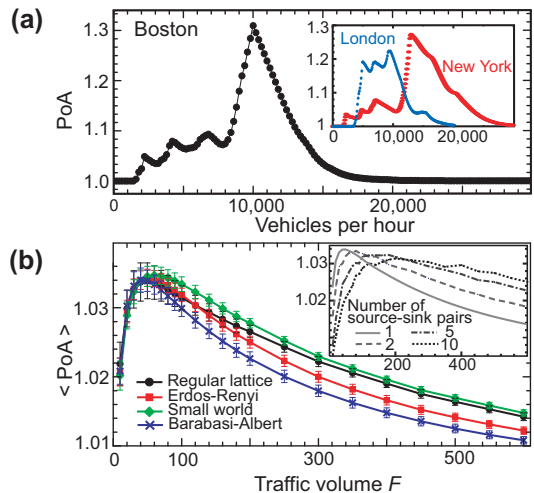}
%\MG(21.07.08)
\caption{(color) The price of anarchy (PoA), as a function of the
traffic volume $F$. (a) In Boston's road network for journeys from
Harvard Square to Boston Common with BPR delays with $\alpha=0.2$,
$\beta=10$. Inset: the PoA in London from Borough to Farringdon,
and in New York from Washington Market Park to Queens Midtown
Tunnel. (b) The PoA in ensembles of model networks with affine
delays.
%MG(20.07.08): This would also be nice to keep, but probably it must go.
%HJ(21.07.08): We can keep it
All networks have 100 nodes and 300 undirected links.
%\HJ(21.07.08): We can keep it
%\MG(20.07.08)
The error bars represent one standard deviation in the
PoA-distribution. Inset: the PoA in regular lattices with multiple
random sources and sinks (``multi-commodity flows'') averaged over
100 to 400 networks. Each pair contributes equally to $F$. }
\label{fig:poa_graph}
\end{figure}

Knowing the PoA is important, but it is even more valuable to
discover a proper method to reduce it. In a road network, one
could charge drivers toll fees to stimulate a more cooperative
behavior, but that strategy has problems of its own.  For example,
one could charge a fee for using each link equal to the ``marginal
cost" $f_{ij} \cdot {l_{ij}}'(f_{ij})$ so that the new Nash flow
becomes equal to the social optimum. Unfortunately, if collected
taxes are not returned to the users, such marginal cost taxes do
not improve the cost of the Nash equilibrium in the case of BPR
delays~\cite{Cole}. However, as we have learned from Fig.~3(b), we
can change the PoA by modifying the underlying network structure.
For instance, closing roads to car traffic is relatively easy to
implement and is, moreover, equally effective for everybody. One
might expect that closing roads only leads to increased
congestion. However, contrary to common intuition, Braess's
paradox suggests that road closures can sometimes reduce travel
delays~\cite{Braess}.

We investigated whether this apparent contradiction occurs in the
road networks of
%MG(20.07.08): Here is a shorter, quite good alternative.
%Boston, London, and New York.
Fig.~2.
%\MG(20.07.08)
In the case of Boston's roads, we set $F$=10,000 between Harvard
Square and Boston Common which is the flow where the PoA reaches
its maximum, i.e., where reducing the travel time is most
desirable.  We then compare the costs of the Nash flow on the
original network with those on networks where one of the 246
streets is closed to traffic. In most cases, the cost increases
when one street is blocked, as intuitively expected. Nonetheless,
there are six connections which, if one is removed, decrease the
delay in the Nash equilibrium, shown as dotted lines in Fig.~2. If
all drivers ideally cooperated to reach the social optimum, these
roads could be helpful; otherwise it is better to close these
%MG(04.08.08): avoiding the repetition of the cities' names.
%streets. Similar results are also found in the road networks of
%London and New York. There are seven links causing Braess's
%paradox in London ($F$=10,000) and twelve in New York
streets. Similar results are also found in the other two networks:
there are seven links causing Braess's paradox in London
($F$=10,000) and twelve in New York
%\MG(04.08.08):
($F$=18,000), see Fig.~2(b) and (c). Of course, the identified
roads may not always be bad because a different set of start and
%MG(04.08.08): avoiding the repetition of ``links causing
%  Braess's paradox''.
%end nodes can change the number and location of links causing
end nodes can change the number and location of links triggering
%\MG(04.08.08)
Braess's paradox. However, their existence under the investigated
conditions suggests that Braess's paradox is more than an academic
curiosity~\cite{Braess,Steinberg} or an anecdote with only sketchy
empirical evidence~\cite{knodel}. Nevertheless, more work is
needed to generalize the presented results, for example for
multiple sources and destinations.  As a first step, we have
calculated the PoA for such multi-commodity flows (Fig.~3(b)
inset).

Braess's paradox exists because the social optimum and the Nash
equilibrium react in different ways to changes in the network.
After a link is closed, the socially optimal travel time must be
at least as long as before. However, there is no a priori reason
why severing a link could not improve the Nash travel time. By the
same argument, adding new links can potentially create more delay
in the Nash equilibrium. Hence, a target for future policies in
transportation networks is to prevent unintended delays caused by,
ironically, well-intentioned new constructions that form a
disadvantageous Nash flow. Because convex costs such as the BPR
function are common in economics, Braess's paradox is presumably
also frequent outside vehicle transportation networks. In fact, we
do not need game theory to find this paradox. It also occurs in
physical networks where equilibrium principles can drive the
network away from optimality. For example, currents in electric
circuits do not always minimize the dissipated energy, but instead
satisfy Kirchhoff's laws. As a consequence, removing wires can
sometimes counter-intuitively increase the conductance
~\cite{Penchina}. Although electrons in a circuit, unlike drivers
in a road network, do not act selfishly, the equilibrium
conditions (Kirchhoff's laws and Wardrop's principle) are in fact
closely related. Further studies of the price of anarchy and
Braess's paradox might therefore lead to significantly improved
flows in a number of important applications.

\begin{acknowledgments}
We thank Eric Smith, Yueyue Fan, D.-H. Kim, and H.-K. Lee for
%MG(04.08.08): It's either ``HY'' or ``H.~Y.''.
%helpful discussion. H.Y acknowledges the motivation from the study
helpful discussion. H.~Y. acknowledges the motivation from the study
%\MG(04.08.08)
group at the NECSI summer school. This work was supported by KOSEF
through the grant No. R17-2007-073-01001-0 and is dedicated to the
memory of Charles VanBoven.
\end{acknowledgments}

\end{document}